\documentclass[aps,pra,showpacs, oneside,preprintnumbers, amsmath,preprint, superscriptaddress, 9pt]{revtex4-1}

\usepackage{amsmath}
\usepackage{amssymb}
\usepackage{graphicx}
\usepackage{multirow}
\usepackage{epstopdf}
\usepackage{booktabs}
\usepackage{ulem}

\usepackage{dcolumn}% Align table columns on decimal point
\usepackage{bm}% bold math
\newcommand{\ignore}[1]{}

\begin{document}

\def\e{\mathcal{E}}

\title{Electromagnetically Induced Transparency (EIT) Amplitude Noise Spectroscopy}

\author{Ben Whitenack}
\affiliation{Lewis \& Clark College, Portland, OR 97219}

\author{Devan Tormey}
\affiliation{University of Southern California, Los Angeles, CA 90007}

\author{Michael Crescimanno}
\affiliation{Department of Physics and Astronomy, Youngstown State
University, Youngstown, OH 44555-2001}

\author{Andrew C. Funk} 
\affiliation{Lewis \& Clark College, Portland, OR 97219}

\author{Shannon O'Leary} 
\affiliation{Lewis \& Clark College, Portland, OR 97219}

\date{\today}

\begin{abstract}
Intensity noise cross-correlation of the polarization eigenstates of light emerging from an atomic vapor cell in the Hanle configuration allows one to perform high resolution spectroscopy with free-running semiconductor lasers.  This shows promise as an inexpensive and simpler approach to magnetometry and timekeeping, and as a probe of dynamics of atomic coherence in warm vapor cells. We report here that varying the post-cell polarization state basis yields intensity noise spectra which reveal details about the prepared atomic state. We advance and test the hypothesis that the observed intensity noise can be explained in terms of an underlying stochastic process in light field amplitudes themselves. Understanding this stochastic process in the light field amplitudes themselves provides an additional test of the three level system model of EIT noise.

\end{abstract}

% note that the PACS classification has been apparently superseded by a n
% new method called "PHySH". Here's the categories I found for that: 
%PHySH: Optics/lasers + Effects of atomic coherence

%PHySH: Atomic/Molecular processes + Magneto-optical spectra

%older pacs code system...depreciated...
%\pacs{42.50.Gy, 32.70.Jz, 34.80.Pa}

\maketitle

\section{INTRODUCTION}
\label{sec:intro}
Electromagnetically induced transparency (EIT)  is a coherent multiphoton effect that results in an optical medium that would normally absorb light becoming ``transparent'' (i.e. being much less absorbing)\cite{EITreview}.  EIT occurs due to quantum interference between the atomic transition pathways that couple to the optical fields.   The optical frequency dependence of the quantum interference required for EIT, makes this process potentially useful for precision spectroscopic applications such as clocks, magnetometers, communication schemes, and quantum computation \cite{Lukinreview,vanderWalScience, duan2001nature, CPTreview, budker,knill_nature_computation}. \par

If the optical fields used for an EIT experiment have narrow frequency spectra ($< 1$ MHz), as is typical in a setup utilizing an external cavity diode laser (ECDL), it is relatively straight-forward to understand EIT and the fluctuations in the light after it has interacted with the medium by using a three level Lambda scheme model which is illustrated in Fig. \ref{setup}b.  In this case we treat the light as classical single mode field and then calculate/measure the relevant properties of the light after exiting the medium.  \cite{PRL1991, McIntyre, rosenbluh,Felinto13,Xiao14b,Rao17}.  Less coherent sources, such as the generic free running semiconductor laser diode used in this experiment, have broad frequency spectra ($\approx $ 80 MHz) and the resulting EIT atom-light interaction is not well modelled in terms of polarization states of a single optical mode. This is also indicated by the low EIT contrast in our experiments. 

EIT intensity noise correlation spectra can be well modelled with a theory based on a reduced number of optical modes and are interesting spectroscopic probes \cite{PRL1991, McIntyre, rosenbluh,Felinto13, Xiao14b, Rao17}. They are understood as examples of the more general process of phase noise to amplitude noise conversion in atomic systems \cite{Camparo, Zoller}.  Here we present  experimental results that test a reduced mode theory for any EIT intensity noise correlations arising from atom-light induced Markov process in the light field amplitudes.  This novel test indicates the robustness of a simple noise model with a classical optical field.   \par

The use of a free-running diode laser to observe the intensity noise correlations in the light from EIT has technological relevance.  A free-running diode laser is far less expensive and more mechanically robust than an ECDL.  The broad optical spectrum of the free-running diode laser is advantageous here as the intensity noise correlations from EIT scale with the laser bandwidth. Furthermore, the use of (phase-)noisy lasers\cite{harvard2} may enjoy metrological advantages over other approaches since they can operate in a regime with good signal to noise but without power broadening\cite{Scully, Martinelli4, Cruz7, Ariunbold10, YanhongPRL, Felinto:13, Moon13, OLeary1,Florez13, Xiao14, Florez14}. 

Previous EIT noise studies indicate that a useful\cite{Scully, harvard2,OLeary1} noise statistic is 
the degree of correlation of the intensity fluctuations captured by
the normalized intensity cross-correlation statistic, $g^{(2)}(0)$:
\begin{equation}
   g^{(2)}(0)= 
\frac{\langle \mathrm{}(\delta
     I_{a})\,\mathrm{}(\delta
     I_{b})\rangle}{\sqrt{\langle(\mathrm{}\delta
       I_{a})^2\rangle\langle(\,\mathrm{}\delta I_{b})^2}\rangle},
\label{eq:g2define}
\end{equation}
where $\delta I_{a(b)}$ is the intensity fluctuation in the selected polarization mode labeled by $a (b)$.  Typically $\delta I_{a}$, $\delta I_{b}$ are the intensity fluctuations of the two propagation eigenstates in the system, and so for a Zeeman EIT system subject to a longitudinal field, these would be the circular polarization states $\sigma_{\pm}$. The numerator of $g^{(2)}(0)$ is the average of the product of the two intensity fluctuations,  and the denominator normalizes the result such that perfect correlation outputs $g^{(2)}(0) = +1$ and perfect anti-correlation yields $g^{(2)}(0) = -1$. While useful, $g^{(2)}(0)$ is clearly insensitive to any relative phase information in the eigenstates, and using just $\sigma_\pm$ basis, is probably too limited for developing a vector magnetometry protocol using EIT noise. \par

In the experiment described below we are able to study $g^{2}(0)$  in an arbitrary basis by correlating the intensity correlations within the standard basis and combining that with a third superposition polarization channel. The additional information provided by this third channel allows us to identify a Markov 
amplitude noise process from which we can derive any intensity noise statistic in any basis. 
Understanding noise in the light field amplitudes themselves enables a more detailed test of the three level system model for EIT noise spectroscopy \cite{harvard2, Ariunbold10, Felinto:13} than 
is possible with intensity noise correlations alone.  \par

The polarization dependence of EIT and EIA signals themselves are relatively well studied experimentally and theoretically \cite{Yudin99, Cartaleva6, Brazhnikov6} and EIT intensity noise spectroscopy in the propagation eigenbasis has been relatively well studied.   \cite{harvard2, Scully,Ariunbold10, Cruz7, OLeary1}. Little has been reported on the polarization dependence of EIT noise spectroscopy. Recent work using restricted sets of pre- \cite{Moon13} and post-cell \cite{Moon14} polarization basis choices could not obtain amplitude noise from experimental data due to the experimental protocols. It is clear in the preceeding references \cite{Moon13, Moon14}that the authors understood the utility of the fact that post-cell polarization selection was not simply creating a linear combination of intensity noise correlations, but algebraically independent ones. \par

In Section \ref{sec:exp} we summarize the experimental setup and protocol used. Section \ref{sec:theory} describes the three level model used in this protocol and how it allows us to frame the experimental results in terms of noise in the underlying field amplitudes. We use the model to semi-quantitatively reproduce the experimentally measured $ g^{(2)}(0)$ as a function of two-photon detuning in different (orthogonal) polarization bases. In Section \ref{sec:results} we test the amplitude noise hypothesis in a model independent way by first inverting experimentally measured intensity correlation data from three channels in a fixed basis to amplitude noise correlations. Then we use those computed amplitude noise correlations to generate intensity noise correlation statistics in other bases and compare directly with experimental measurements carried out in those other bases.

\section{Experimental Setup}
\label{sec:exp}

\begin{figure}
\includegraphics[width=\linewidth]{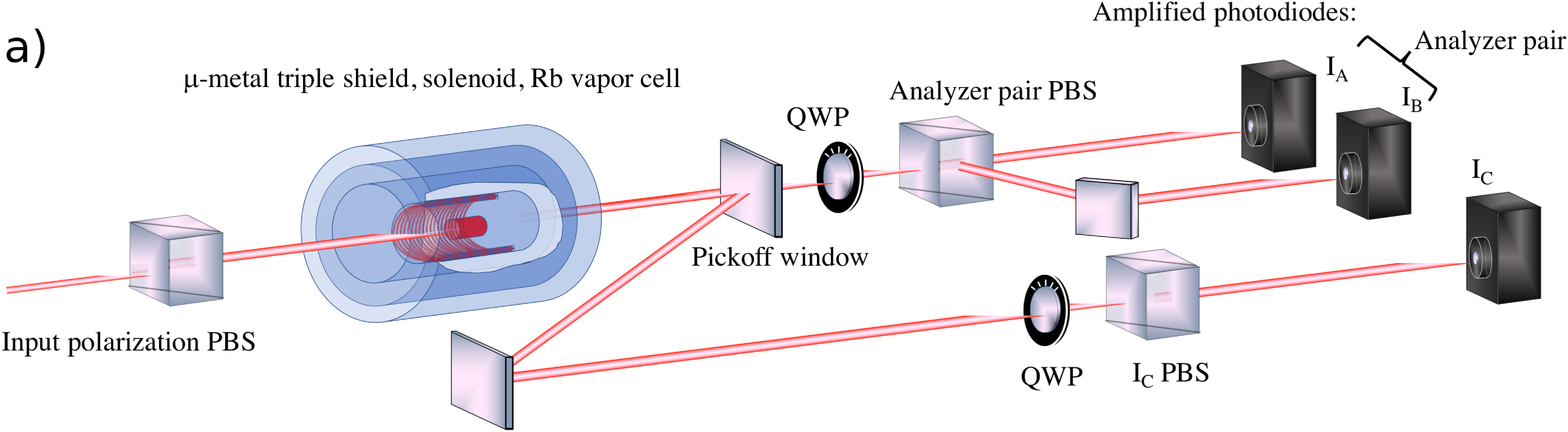}
\includegraphics[width=\linewidth]{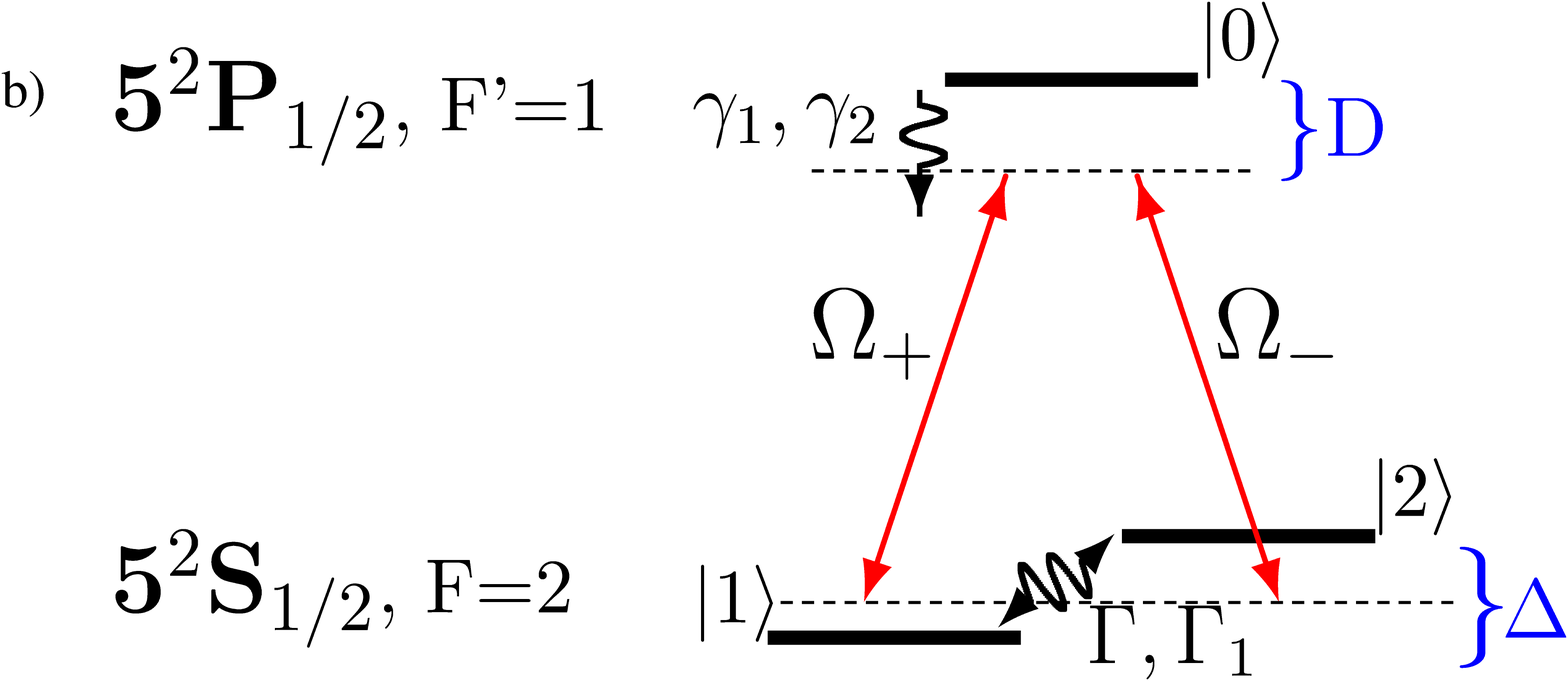}
\caption{(color online) (a) Experimental setup and (b) the 3-level diagram of the minimal atomic quantum optics model of EIT.} 
\label{setup}
\end{figure}
 
The experimental setup is typical for the Zeeman EIT noise spectroscopy and is shown in Fig. \ref{setup}.  The EIT state is realized in a warm ($44^\circ$C) enriched $^{87}$Rb buffer gas vapor cell (length = 8 cm), buffered with 10 Torr of Neon and Argon (5 Torr Ne + 5 Torr Ar )
by a linearly polarized beam from a free-running diode laser sent through the length of the cell.  The linearly polarized input field (in our experiment, horizontal) has equal components $\sigma_+$ and $\sigma_-$ (right and left circularly polarized fields) of fixed phase. This light pumps atomic coherences between degenerate ground state Zeeman sub-levels, as in Hanle EIT.  The degeneracy of the Zeeman sub-levels can be split by an applied longitudinal magnetic field.   The amount of splitting $\Delta$ between the Zeeman sub-levels is determined by the strength of the applied magnetic field, and subsequently we refer to it as a the two-photon detuning. 
To ensure that a reproducible and stable splitting $\Delta$ is only a function of current applied to the solenoid around the vapor cell, it is  
magnetically shielded from the surrounding environment by three nested layers of $\mu$-metal magnetic shielding.  (see \cite{OLeary1} for more experimental details.) \par

The laser field is generated with a  795 nm free-running diode laser which is tuned via temperature and current modulation to the F=2 $\rightarrow$ F'=1 hyperfine transition of the ${}^{87}$Rb  D1 line.  The diode laser's mean frequency is stabilized with a ``loose-lock'' to this hyperfine transition by an analog feedback circuit whose control signal is the saturated absorption signal from another rubidium cell.  The laser's free-running linewidth of $\sim 80$ MHz is unchanged by the loose-lock, which simply prevents long term laser frequency drift; there is no other frequency stabilization of the laser (e.g. no grating feedback - internal or external). The lock used in all the data collected and described here resulted in a modest negative ({\it i.e.} red) one-photon detuning of about 10 MHz.   The large spectral bandwidth of such a ``noisy" laser is desirable for EIT noise correlation studies. The large spectral bandwidth allows one to effectively probe many two photon pathways simultaneously.  The residual intensity noise of the laser was measured to $\sim0.2$\%, and this small amount of residual intensity noise does not affect the utility of the spectroscopic signal.  \cite{Xiao14b,Florez13, zhang1995} .  

Upon exciting the atomic vapor cell, the light is split into  three different polarization components.
For the two light fields from what we call the analyzer pair behind the PBS after the quarter wave plate,  we compute the $g^{(2)}(0)$ using Eq.(\ref{eq:g2define}). 
Referring to Fig. (\ref{setup})a, most of the light ($> 90\%$) is passed through a quarter wave-plate (QWP) before being split by a polarizing beam splitter (PBS) into linearly horizontal and vertical polarization components of what we refer to as the analyzer pair.  When the QWP is aligned at 45$^\circ$ to the horizontal, one can view the QWP as transforming the right and left circular polarization to linear horizontal ($I_a$) and vertical ($I_b$) polarization intensities. 
\par
A small portion of the light ($\approx 8 \%$ for the third channel) is split off before the analyzer pair's quarter-wave plate via a glass window.  The window is aligned at near normal incidence to leave the polarization state of both the transmitted and reflected light nearly unchanged. The polarization state incident to the analyzer pair optics is changed less than 1\% by the pick-off window and all the optics prior to the analyzer pair.  Similar to the analyzer pair light fields, the third channel light is passed through a QWP and a linear polarizer so that the intensity of light incident on its photodiode corresponds to 
a fixed, algebraically independent amplitude combination (throughout this experiment, its QWP axis was rotated an additional 10$^\circ$ with respect to the orientation of the QWP before the analyzer pair that would have led to $\sigma_\pm$ there).   \par

Selecting different polarization bases post-cell by rotating the quarter wave plate before the analyzer pair sends different linear combinations of the (assumed) underlying circular polarization field amplitudes to each of the channels $a$ and $b$. By simultaneously recording all three intensities, $a$ and $b$ and the third channel, we are able to determine all intensity noise correlations. \par

Each light intensity is measured on identical amplified silicon photodiodes whose output voltages are simultaneously digitized by a National Instruments 9223 and recorded  in 4mS windows at a sampling rate of 1 MHz. This is done for different two-photon detunings (generated by the current in the solenoid). All intensity noise correlations are computed numerically from the set of measured intensities.   The amplified photodiode dark current noise power was spectrally flat (DC $\rightarrow$ 10 MHz) and varied between 2 to 12dB below the optical field intensity noise of interest.  The measured intensity noise of each polarization channel has contributions from the conversion of phase noise to intensity noise due to the interaction of the coherently prepared medium with the light, common-mode technical noise, and uncorrelated technical noise.  \par

The effect of common-mode technical noise is to increase the measured intensity noise correlations between the photodiodes and this increases $g^{(2)}(0)$ from its expected value.  The effect of uncorrelated technical noise in the optical fields on the other hand is to decrease the measured intensity noise correlations and thus reduce $|g^{(2)}(0)|$ from its expected value.  The distortions in the $g^{(2)}(0)$ statistic caused by the common-mode and uncorrelated technical noise can be significantly reduced in the experimental data by appropriate Fourier filtering of the raw data.  We collect and correlate data captured at an analyzer wave plate setting for which the intensity noise in the third channel should be perfectly correlated with the noise in the analyzer pair's $\sigma_+$ channel.  The expected normalized intensity cross correlation, $g^{(2)}(0)$ between the third channel and the $\sigma_{+}$ channel of the analyzer pair is thus $+1$ for all two photon detunings (because they are measuring the noise in the same optical polarization), with any reduction in $g^{(2)}(0)$ due to extraneous (non-atomic) technical noise.  Similarly, $g^{(2)}(0)$ between the third channel and the $\sigma_{-}$ channel as a function of two-photon detuning should match the $g^{(2)}(0)$ between the $\sigma_{+}$ and $\sigma_{-}$ channels.  However any common-mode technical noise increases the noise correlation  between the channels and their $g^{(2)}(0)$ increases from $-1$ when the two photon detuning $\Delta$ is in the range where the noise in the two polarizations it anti-correlated (i.e. $ 500$ Hz $  < | \Delta | < 1$ MHz).  
A band-pass Fourier filter of the noise data recovers this expected behavior, enabling us to determine an electronic frequency window that greatly reduces the contribution to $g^{(2)}(0)$ from both the uncorrelated and the common mode technical noise (see also \cite{Jeong17}). \par

\section{Theory}
\label{sec:theory}

A three level atomic model quantitatively captures the noise properties of EIT amplitude and 
intensity correlations. (see, for example, Refs. \cite{harvard2, OLeary1}). The  model
consists of a three level $\Lambda$ system (ground states $| 1 \rangle$ and $| 2 \rangle$, and excited state $| 0 \rangle$, see Fig. (\ref{setup})b) with the output light's amplitudes, in magnitude and phase, encoding different off-diagonal density matrix elements. 
The intensity cross-correlation statistic in the optically thin cell limit is computed by using the appropriate density matrix elements of the static solution 
to represent the slow-evolving parts of the density matrix. These include the ground state populations and coherences. To then determine the light field amplitudes post cell, for fixed slow-evolving parts of the density matrix one continually updates the fast-evolving parts (the parts of the density matrix involving the excited states for example) while ensemble averaging over a flat distribution of one photon detunings (a process meant to model the laser diode's phase noise, assumed spectrally broader than the transition's intrinsic one-photon width). With just the intensity of the two circular polarization channels $\sigma_+$ and $\sigma_-$ it is not possible to extract the underlying amplitude noise spectra. Thus a third intensity channel is added that measures light in a fixed, independent linear combination of amplitudes that make up  $\sigma_+$ and $\sigma_-$. Refer to the intensity measured in this third channel as $I_c$.

The intensity cross correlation function $g^{(2)}(0)$ (as defined above) quantifies only the the intensity noise correlations between two optical polarization modes.  In general, there exist additional noise correlations not necessarily captured by $g^{(2)}(0)$.  For the case of optical polarization the full set of second order intensity correlations can be written in terms of six amplitude two-point functions $<a^2>, <b_1^2>, <b_2^2>, <a b_1>, <a b_2>$, and $<b_1 b_2> $; where $a$ is the amplitude of one optical polarization mode and $b_1(b_2)$ is the real (imaginary) part of the orthogonal polarization mode.  All information about the optical polarization is invariant to global phase shifts, so without loss of generality $a$ is real. 

We do not assume Gaussian statistics for these amplitudes components, but the six two-point functions above are sufficient (in leading order) to compute the intensity noise correlations for any post-cell beam pair. Both our experimental data itself and separately the three level 
model indicate that the three point noise amplitude functions (and higher correlations) are not significantly smaller than the two point amplitude noise correlations, but we do not address this further here. Since we do not need the higher order correlations for the leading order intensity correlations, we fix the post cell polarization basis and compute the six independent two point functions as a function of two-photon 
detuning $\Delta$. Using the measured intensity noise correlations $<\delta I_a \delta I_a> , <\delta I_b \delta I_b>, <\delta I_c \delta I_c>$ , $<\delta I_a \delta I_b> , <\delta I_b \delta I_c>, <\delta I_a \delta I_c>$ and assuming the light is expressible in terms of the two amplitudes $a$ and $b$, we compute the
six independent amplitude correlations as a function of two-photon detuning. 

We also invert this process, using the six amplitude correlations measured in one of the polarization bases to construct the expected $g^{(2)}(0)$ cross-correlation statistic for any other polarization basis. 
Finally, we independently experimentally measure those intensity correlations by choosing different polarization state bases (via rotating the wave plate before the analyzer pair) and compare those data with the amplitude noise reconstructed expectations, resulting in a  semi-quantitative test of treating the noise as a single mode Markov process. 
These tests show how well any intensity statistic of light emerging from a free running laser diode and subsequently processed by an EIT medium can be quantitatively constructed in terms of a stationary Markov process in a single pair of light field amplitudes.

\section{EXPERIMENTAL RESULTS AND DISCUSSION}
\label{sec:results}

\begin{figure}
\includegraphics[width=4.0in, height=2.3in]{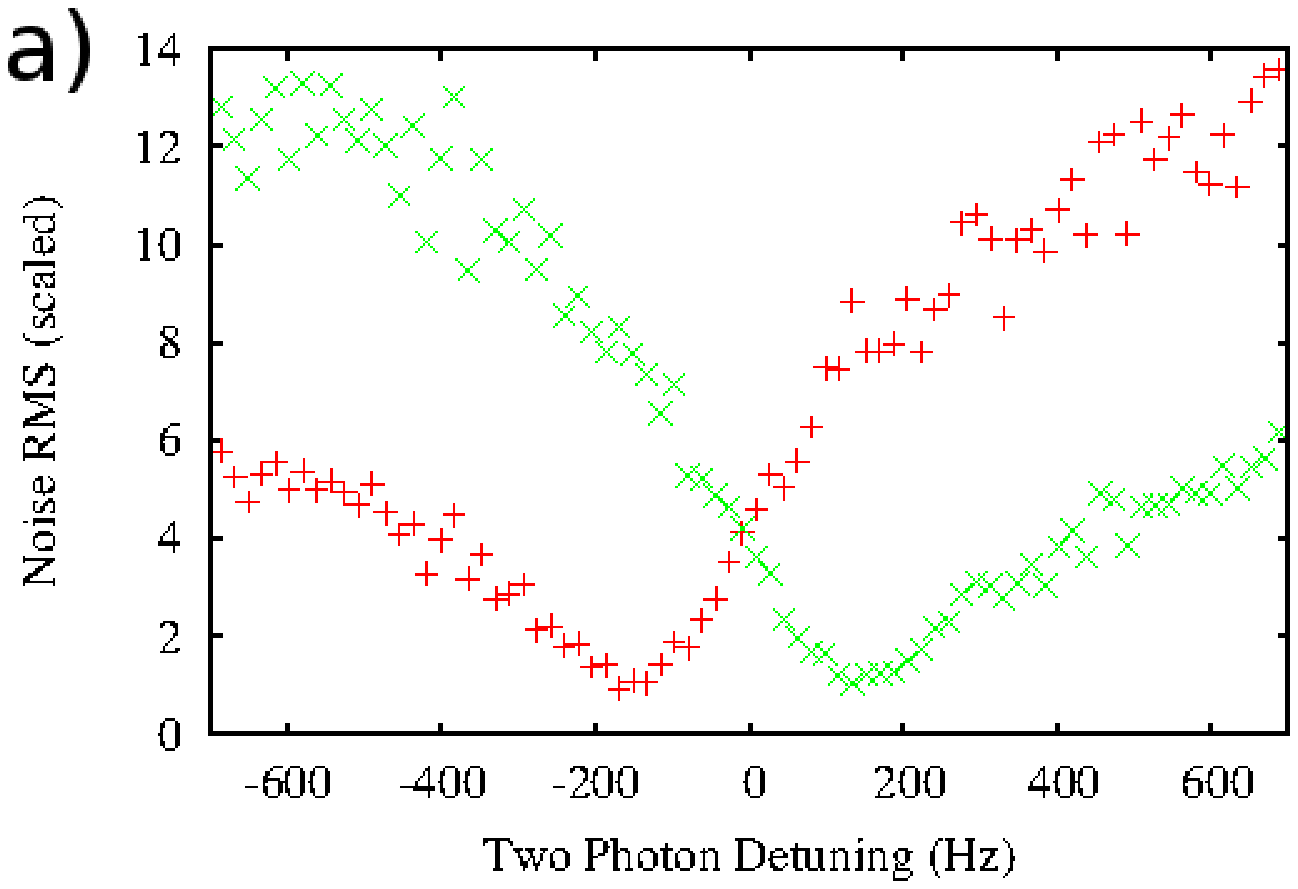}
\includegraphics[width=4.0in, height=2.3in]{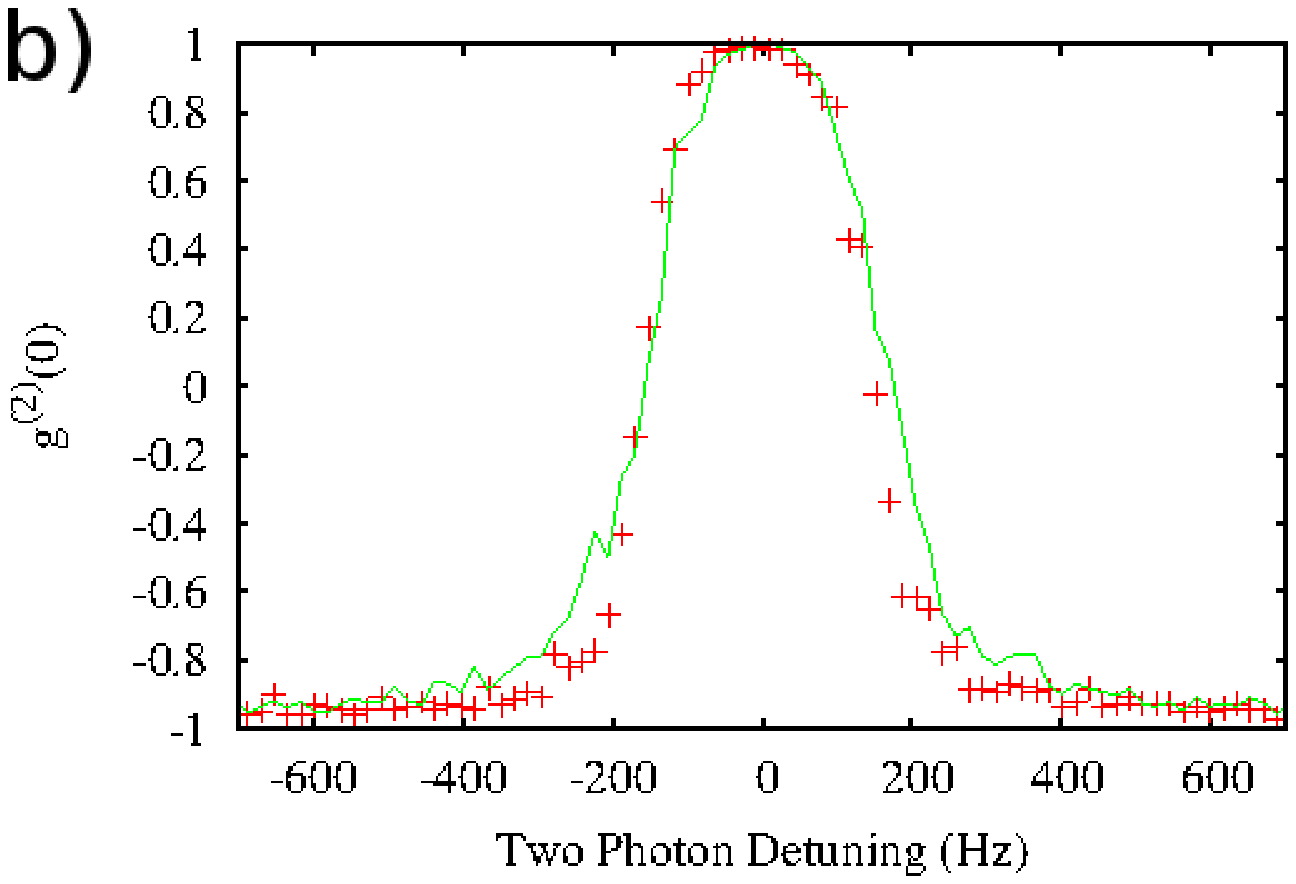}
\includegraphics[width=4.0in, height=2.3in]{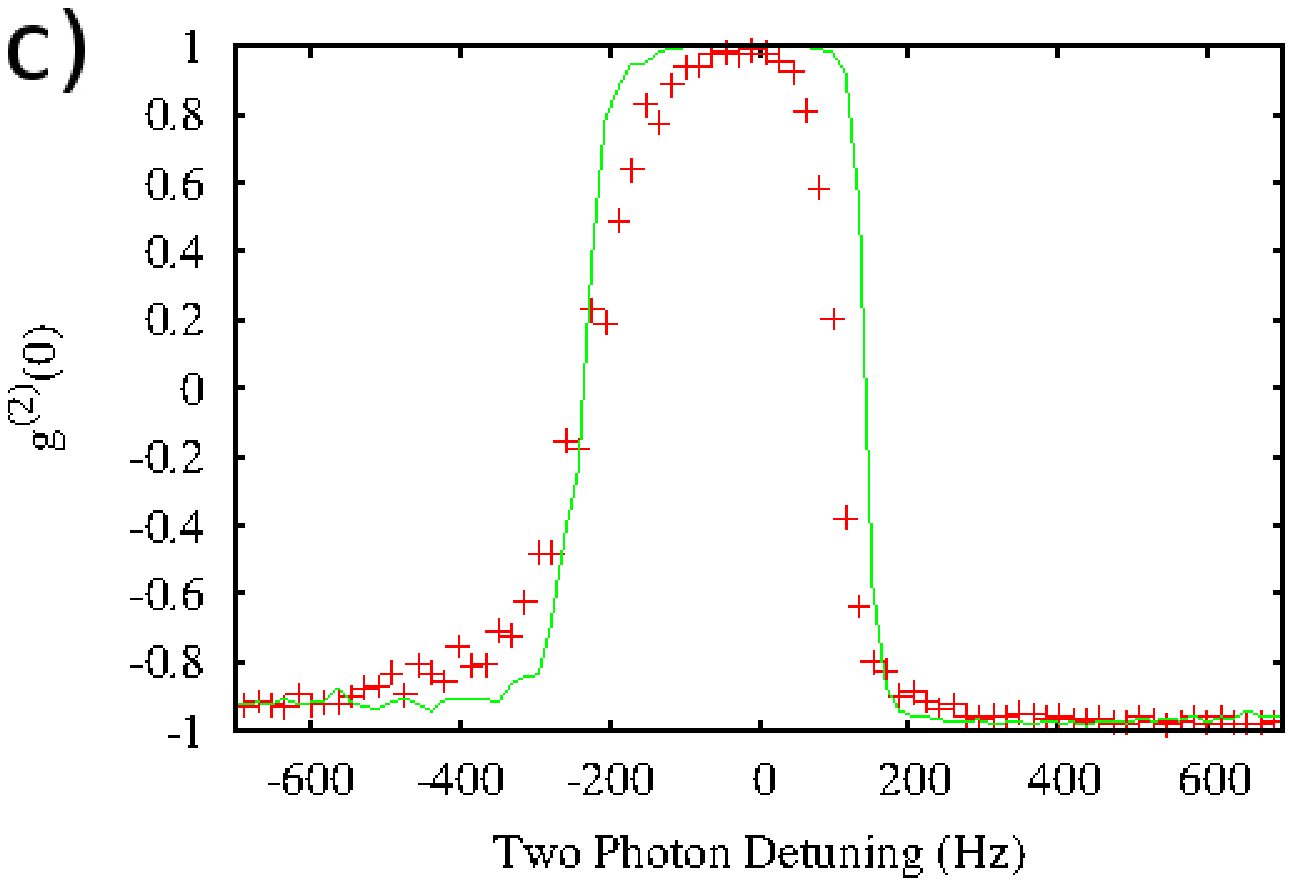}
\caption{(color online) (a) Experimental RMS noise in each of the analyzer ports 
as a function of the two-photon detuning with the post-cell phase plate held at 45$^\circ$  ($\sigma_+$ in Red '+'s, and $\sigma_-$ in green 'x's). (b) The cross-correlation statistic, $g^{(2)}(0)$ versus detuning measured (points) and theory (line) for the cross-correlation statistic for the post-cell phase plate held at 45$^\circ$ and (c) Experiment (points) and theory (line)  for the post-cell quarter wave plate held at 60$^\circ$. The lack of smoothness of the theory curves is due to finite sample size for each detuning (here chosen to match that of the experimental data)} 
\label{noise_traces}
\end{figure}

When the post-cell quarter wave plate is aligned at $45^{\circ}$, the measured noise correlation is between the $\sigma_+$ and $\sigma_{-}$,  the propagation eigenstates for the EIT process. In this case, the contrast of $g^{(2)}(0)$ generally increases with power and its central feature eventually broadens. All the data shown here was taken at a optical power and beam diameter which corresponds to an intensity just beyond the beginning of the power broadening regime (\cite{OLeary1}). Rotating the post-cell quarter wave plate before the analyzer pair from its nominally 45$^{\circ}$ orientation with respect to the input polarization,  we record systematic changes in the subsequently measured intensity correlation noise spectra in the analyzer detector pair. As a function of the two-photon detuning $\Delta$, typical RMS noise traces in each port of the analyzer are shown in  
Fig.\ \ref{noise_traces}a  and typical $g^{(2)}(0)$ from these data (points) and associated theory (lines) are shown in Fig.\ \ref{noise_traces}b and c. We summarize our comparison between experiment and theory for EIT noise after a polarization basis change as follows: 

% PLotting instructions for (a) 
%set datafile separator "," 
%set terminal post color enhanced portrait "Times-Roman" 28 size 3.5,2.6
%set ylabel "Noise RMS (scaled)"
%set xlabel "Two Photon Detuning (Hz)" 
% unset key
% set xtics scale 0.5
% set ytics scale 0.5  
% set border lw 1.5
% plot [-700:700] "noise_filtered_RMS0.csv" using ($1*1000):($2*1e4), "noise_filtered_RMS0.csv" using ($1*1000):($3*1e4) 
% set output "outplot.ps"
% replot

% plotting instructions for (b)
% set ylabel "g^{(2)}(0)"
%plot [-700:700] "noise_filtered_RMS0.csv" using ($1*1000):8, "noise_filtered_RMS_sim0.csv" using (-$1*1000):8 w l lt 1 lc rgb 'green'

% plot instructions for (c) 
%plot [-700:700] "noise_filtered_RMS15.csv" using ($1*1000):8, "noise_filtered_RMS_sim15.csv" using (-$1*1000):8 w l lt 1 lc rgb 'green'

\begin{figure}
\includegraphics[width=\linewidth]{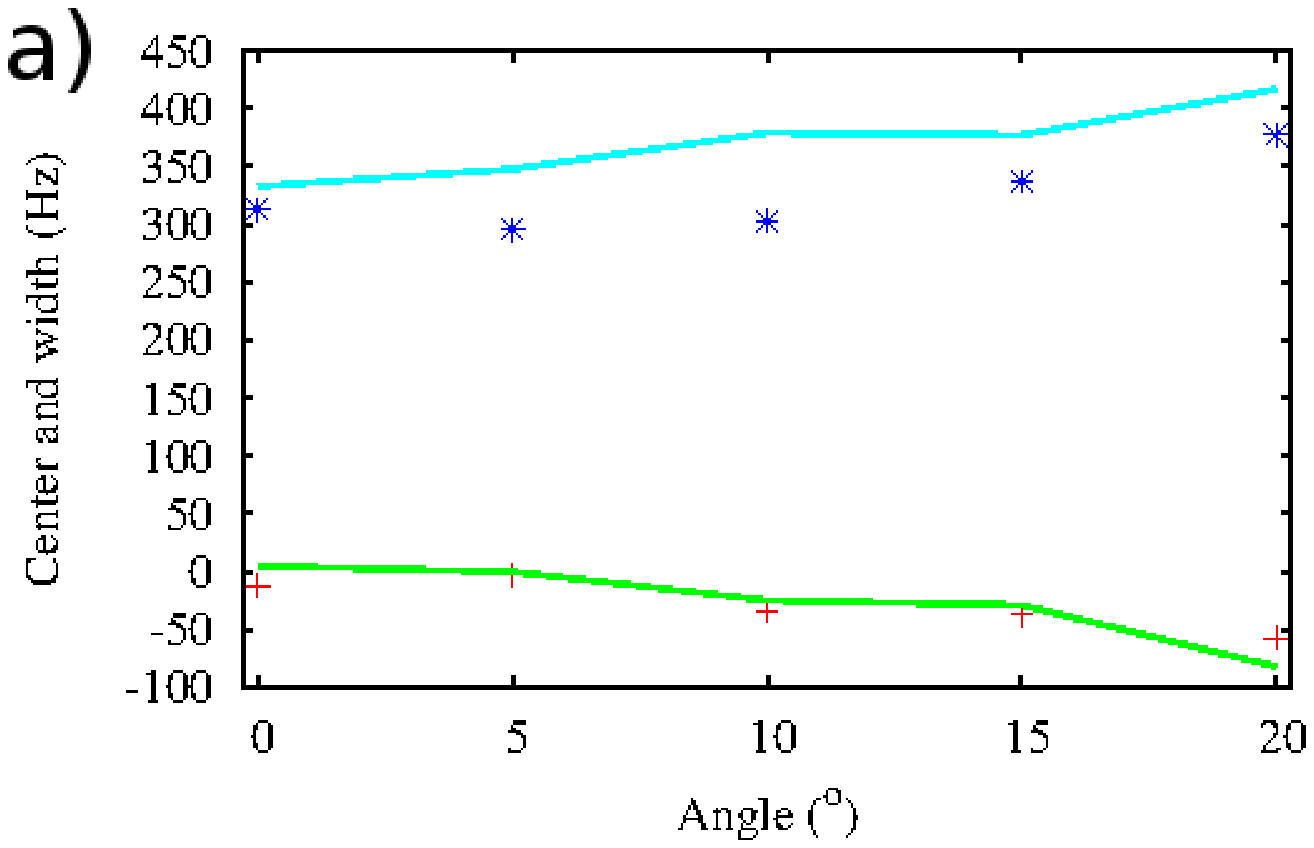}
\includegraphics[width=\linewidth]{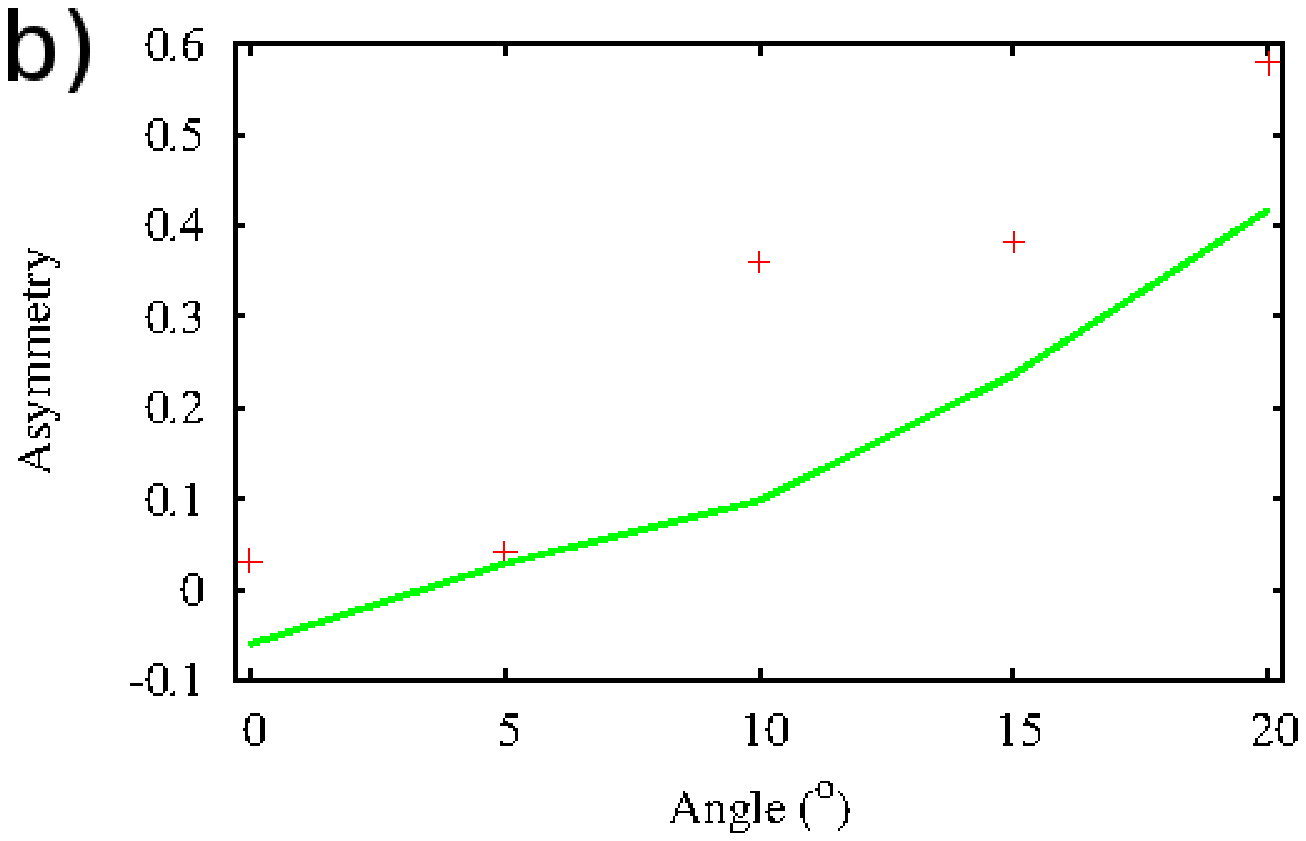}
\caption{(color online) (a) Shift of the $g^{(2)}(0)$ maxima, experiment (red "+"'s') and theory (green line) with wave plate angle beyond 45$^\circ$ degrees. Also shown is the FWHM, experiment (blue stars) and theory (cyan line). (b) The asymmetry $\upsilon$ as a function of the wave plate angle beyond 45$^\circ$ degrees.}
\label{shifts_asym}
\end{figure}

\textbf{1) Shift:} For rotations of the QWP beyond 45$^\circ$,  the metrologically relevant central peak of $g^{(2)}(0)$  is no longer at zero two-photon detuning. The direction and magnitude of the shift is well accounted for in the three level atomic model, \cite{harvard2, OLeary1}. For rotations of the post-cell quarter wave plate beyond its nominal 45$^\circ$ orientation we show in Fig.\ \ref{shifts_asym}a the shift and width of the $g^{(2)}(0)$ as a function of 
additional rotation beyond 45$^\circ$. 
Although we have not included it here, the shift of the peak with power at fixed angle is also well modeled with the theory model. Also, as expected, the signs of the shift and the asymmetry (described below) flip with the sign of the one photon detuning.

\textbf{2) Asymmetry:} When the QWP is at the nominal 45$^\circ$ rotation, $g^{(2)}(0)$  is symmetric in the two-photon detuning. For rotations of the QWP beyond 45$^\circ$,  $g^{(2)}(0)$ 
becomes asymmetric, and so we define the asymmetry $\upsilon = {{W_+-W_-}\over{W_++W_-}}$ where
the unsigned half-widths $W_\pm$ are the magnitudes of the detunings at which the $g^{(2)}(0)$ crosses zero. Fig.\ \ref{shifts_asym}b is a plot of $\upsilon$ for both experiment (dots) and theory (lines). The theory further  indicates that the sign of the asymmetry should flip with the sign of the angle of the quarter wave plate orientation beyond 45$^\circ$ degrees, an effect that was experimentally verified but has not been separately quantified. The fit parameters used for this comparison between theory and experiment are the total ground state decoherence rate and the optical power, fixed by data from only the 45$^\circ$ quarter wave plate orientation. 

% these figures made from "lineshape_data" file, pretty straightforward since there are only like 5 points! Use the formatting from above Figure 2! 
%for figure (a) : 
%set ylabel "Center and width (Hz)"
%set xlabel "Angle (^o)"
%plot [-.3:20.3] "temp" using 1:3, "temp" using 1:6 w l lt 1 lw 3 lc rgb "green", "temp" using 1:($4-$2), "temp" using 1:($7-$5) w l lt 1 lw 3 lc rgb "cyan"
% for figure (b): 
%set ylabel "Asymmetry"
%plot [-.3:20.3] "temp" using 1:(-($2+$4)/($4-$2)), "temp" using 1:(-($5+$7)/($7-$5)) w l lt 1 lw 3 lc rgb "green"

Next we use a single simultaneous data set consisting of the analyzer pair {\bf and} a third channel at a fixed different quarter wave plate rotation to construct an expected $g^{(2)}(0)$.
We simultaneously collected data using the third channel whose quarter wave plate was rotated 10$^\circ$ 
beyond that of the quarter wave plate in front of the analyzer pair.  Combining the simultaneous data
from each of the three channels, we numerically computed the six amplitude noise correlation two-point functions $<a^2>, <b_1^2>, <b_2^2>, <a b_1>, <ab_2>$, and $<b_1 b_2> $. 
There were two fit parameters used to compute these correlations; one that fixed the relative intensity ratio of between the third channel and the analyzer pair (close to 8\%, as measured) and the other an intensity offset applied to the signal level of the third channel alone. 
We use the two-point amplitude correlations to estimate the intensity $g^{(2)}(0)$ in a different post-cell polarization basis from rotating the wave plate in front of the analyzer pair by a fixed amount. 

% to generate each part of this frame: 
% set ylabel "g^{(2)}(0)"
% set xlabel "Two-Photon Detuning (Hz)" 
% unset key
%plot [-700:700] "noise_filtered_RMS5.csv" using ($1*1000):8, "fixed_scale5.csv" using ($1*1000):9  w l lt 1 lw 3 lc rgb "green"
%plot [-700:700] "noise_filtered_RMS10.csv" using ($1*1000):8, "fixed_scale10.csv" using ($1*1000):9  w l lt 1 lw 3 lc rgb "green"
%plot [-700:700] "noise_filtered_RMS15.csv" using ($1*1000):8, "fixed_scale15.csv" using ($1*1000):9  w l lt 1 lw 3 lc rgb "green"
%plot [-700:700] "noise_filtered_RMS20.csv" using ($1*1000):8, "fixed_scale20.csv" using ($1*1000):9  w l lt 1 lw 3 lc rgb "green"

\begin{figure}
\includegraphics[width=\linewidth]{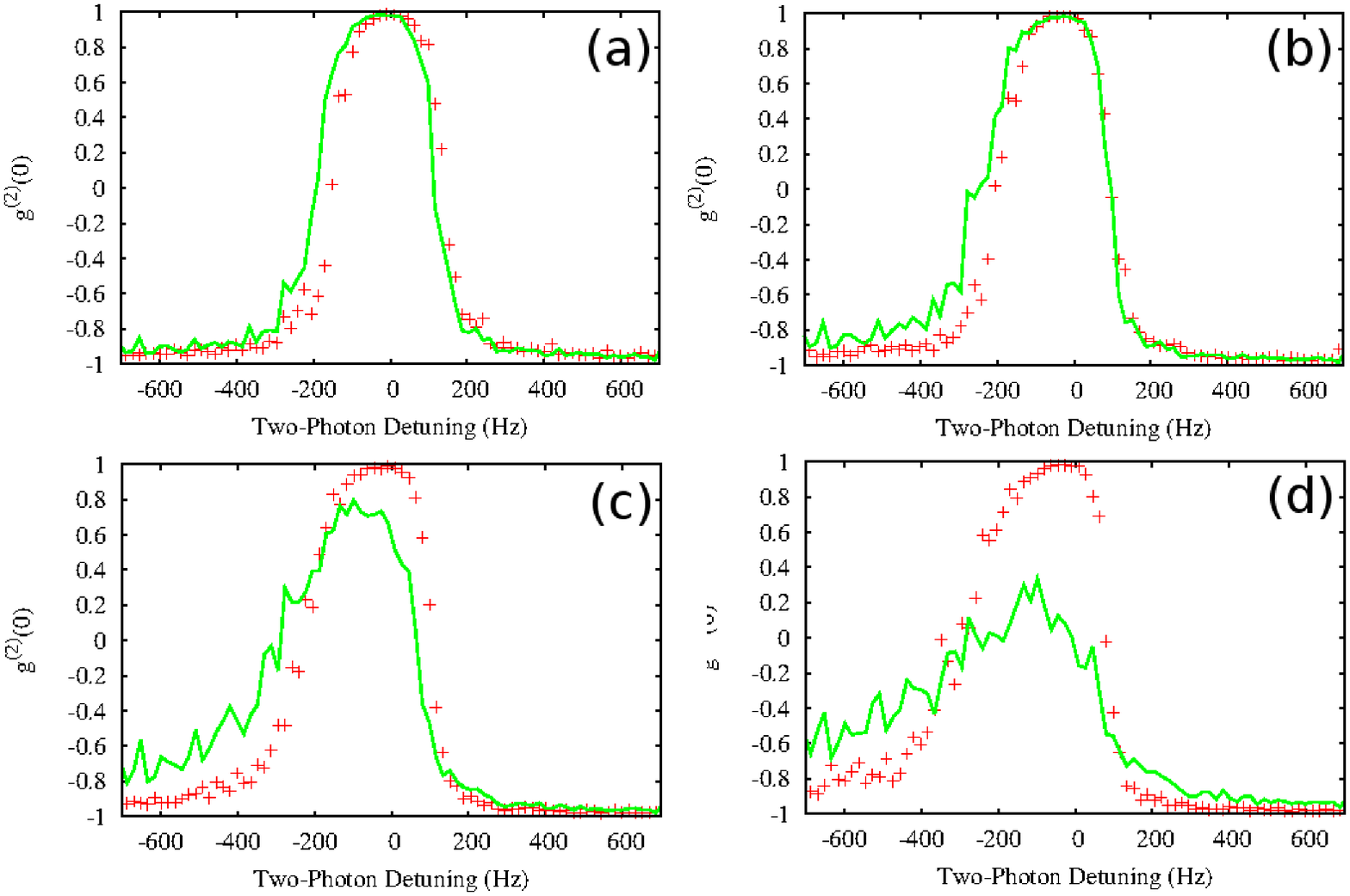}
\caption{(color online) Measured $g^{(2)}(0)$ EIT intensity correlation curves (in green, 'x's) and their reconstruction using amplitude noise analysis (in red, as "+"s), as a function of the two-photon detuning at different post cell quarter wave plate orientations. All angles are displacements from the nominal 45$^\circ$ orientation. (a) 5 degrees additional rotation, (b) rotated an additional 10 degrees, (c) additional 15 degrees, (d) additional 20 degrees rotation.} 
\label{recon_quad_panel}
\end{figure}

In Fig.\ \ref{recon_quad_panel} we plot this computed $g^{(2)}(0)$ estimate along with measured $g^{(2)}(0)$
data (recorded separately) in the same polarization basis for four different rotations of that quarter wave plate. This computed $g^{(2)}(0)$ predicts the overall shift and asymmetry. The computed $g^{(2)}(0)$ degrade at larger quarter-wave plate angle, as shown not only by their differences from the measured curves, but also by the fact that they become more noisy. We have verified numerically that additional uncorrelated noise tends to make data as seen in Fig.\ \ref{recon_quad_panel}a,b more 
like Fig.\ \ref{recon_quad_panel}c,d.  Up to questions regarding the precise cause of the reconstruction's degradation at large angle, the measured intensity noise correlations  compared with the amplitude noise reconstructed versions indicate the validity of the reduced mode, three level model for intensity (and amplitude) noise of EIT driven by a free-running laser diode. 

We have not included theory fits to the data only to simplify the presentation and focus this present paper on question of the utility, economy and completeness of deriving the observed intensity noise structure of the light emerging from this coherently prepared medium in terms of amplitude noise correlations. Likewise, we did not display theory noise RMS traces at various quarter wave plate angles (as in  Fig.\ \ref{noise_traces}a), though they are quite similar to the figure shown.

\section{Conclusions and Acknowledgements}
\label{sec:conclusion}

We have tested an approach to understanding intensity noise spectra from an EIT system as a Markov process in the one-photon detunings of a single input light field amplitude. Our results indicate that
one can reliably compute EIT intensity noise correlations in an arbitrary polarization basis by identifying a set of amplitude noise correlations in a given polarization basis. Since vector magnetometers detect the transverse components of a magnetic field through changes in the 
ellipticity of the light fields, the tests performed here suggest a new way of using EIT noise protocols in vector atomic vapor magnetometry and  indicate potential utility and robustness of using noise spectroscopy in device applications such as atomic clocks \cite{kitching01a} and magnetometers \cite{CPT_magnetometer}.

We are grateful to I. Novikova, Y. Xiao, D.\ Phillips and  R.\ Walsworth for 
discussions and equipment use.  Much of this work was supported under NSF grant number PHY-1506499 (SO, AF, BW). and MC acknowledges support under NSF grant number DMR-1609077.

\bibliography{references}
\end{document}